\documentclass{aa}  

\usepackage{bm}

\usepackage{comment}

\def \bA {{\bf A}}
\def \bB {{\bf B}}

\def \bj {{\bf j}}

\def \br {{\bf r}}

\def \del {{\bm \nabla}}
\def \half {\textstyle{\frac{1}{2}}}

\def \curl {{\bm \del} \times}
\def \p {\partial}

\def \. {\cdot}

\def \be {\begin{equation}}
\def \ee {\end{equation}}

\usepackage{graphicx}

\usepackage[dvipsnames]{xcolor}

\usepackage{soul}

\usepackage{txfonts}
\begin{document}

   \title{Chromospheric and coronal heating and jet acceleration due to reconnection driven by flux cancellation}

   \subtitle{I. At a three-dimensional current sheet}
   \titlerunning{Heating by Flux Cancellation}

   \author{E.R. Priest
          \inst{1}
          \and
          P. Syntelis\inst{1}
          }

   \institute{School of Mathematics and Statistics, University of St. Andrews, Fife KY16 9SS, Scotland, UK\\
              \email{psyntelis@gmail.com}
             }

   \date{Received ; accepted }

 
  \abstract
   {The recent discovery of much greater magnetic flux cancellation taking place at the photosphere than previously realised has led us in our previous works to suggest magnetic reconnection driven by flux cancellation as  the cause of a wide range of dynamic phenomena, including jets of various kinds and solar atmospheric heating.  }
   {Previously, the theory considered  energy release at a two-dimensional current sheet. Here we develop the theory further by extending it to an axisymmetric current sheet in three dimensions without resorting to complex variable theory. }
   {We analytically study reconnection and treat the current sheet as a three-dimensional structure. We apply the theory to the cancellation of two fragments of equal but opposite flux that approach each another and are located in an overlying horizontal magnetic field.}
   {The energy release occurs in two phases. During Phase 1, a separator is formed and reconnection is driven at it as it rises to a maximum height and then moves back down to the photosphere, heating the plasma and accelerating a plasma jet as it does so. 
   During Phase 2 the fluxes cancel in the photosphere and accelerate a mixture of cool and hot plasma upwards.}
   {}


  \maketitle
%

\section{Introduction} 
\label{sec1}
Observations of the photospheric magnetic field at a resolution of 0.15 arcsec from the Sunrise balloon \citep{solanki10a,solanki17b} have shown that the rate of magnetic flux emergence and cancellation is an order of magnitude higher than previously thought \citep{smitha17}. In addition, coronal loops have been found to be invariably rooted in mixed polarity fragments that are cancelling at a rate of typically 10$^{15}$ Mx sec$^{-1}$, and the loops brighten when photospheric magnetic flux cancels \citep{tiwari14,chitta17b,chitta18,huang18}. 
Flux cancellation has also been associated with the acceleration of jets on a variety of scales \citep{sterling15,sterling16a,sterling17,panesar18,samanta19,panesar20}.

The Sunrise observations led \citet*{priest18} to propose a `cancellation nanoflare model' for heating the chromosphere and corona, not just X-ray bright points, for which flux cancellation had previously been suggested as a mechanism \citep{priest94b,parnell95}.
The model was supported and extended by numerical simulations, which also showed that various kinds of reconnection-driven jets can form during flux cancellation \citep{syntelis19,syntelis20}.

\citet*{priest18} considered in particular a model in which two magnetic fragments of flux $F$ and $-F$ approach each other and cancel in an overlying horizontal magnetic field of strength $B_0$.  Initially, when far apart, the magnetic sources are not connected, but separator reconnection \citep{priest96a,longcope96a,galsgaard96,parnell08a} starts to occur when the half-separation (d) of the sources becomes less than the flux interaction distance \citep{longcope98},
\begin{equation}
d_0=\left(\frac{F}{\pi B_0}\right)^{1/2}.
\label{eq1}
\end{equation}
As $d$ decreases and the flux sources approach each other, the separator rises to a maximum height of $0.6 d_0$ and then falls to the solar surface as the sources come into contact.  The maximum height at which reconnection occurs therefore depends on the flux ($F$) and field strength ($B_0$) through the parameter $d_0$, and it may be located in the chromosphere, transition region, or corona.  Numerical experiments studying this scenario have since reinforced the validity of the model \citep{syntelis19,peter19,syntelis20}, and as have recent observations \citep{park20}.

In this series of papers, we plan to develop the basic theory in several directions. Here we  remedy a deficiency in the theory, namely, that the properties of the current sheet have so far usually been analysed using complex variable theory,  according to which the input magnetic field at the entrance to a current sheet of length $L$ is related to the sheet length ($L$) and the field gradient ($k$) near the initial null point or separator by
\begin{equation}
B_i=\half kL, 
\label{eq2}
\end{equation}
which is a key result used in the theory.

Thus, since complex variable theory applies only in two dimensions (2D), the theory so far is valid only in 2D. In the present paper, we therefore develop a corresponding theory for a current sheet in three dimensions (3D), in particular for an axisymmetric current sheet, and derive a generalisation of the result in Eq. (\ref{eq2}).  We also apply this new result to the case of reconnection driven by the approach of equal and opposite flux sources in an overlying uniform horizontal magnetic field studied in \citet*{priest18}.  First of all, we consider a 2D current sheet and show how the expression for the magnetic field around it can be obtained by the new method without using complex variable theory (Sect. \ref{sec2.2}). Then we generalise this method to the field of a 3D axisymmetric current sheet (Sect. \ref{sec3.2}), and finally we apply it to the creation of such a sheet by the flux cancellation of two flux sources (Sect. \ref{sec4}).

\begin{figure}[h]
    \centering  
    \includegraphics[width=\columnwidth]{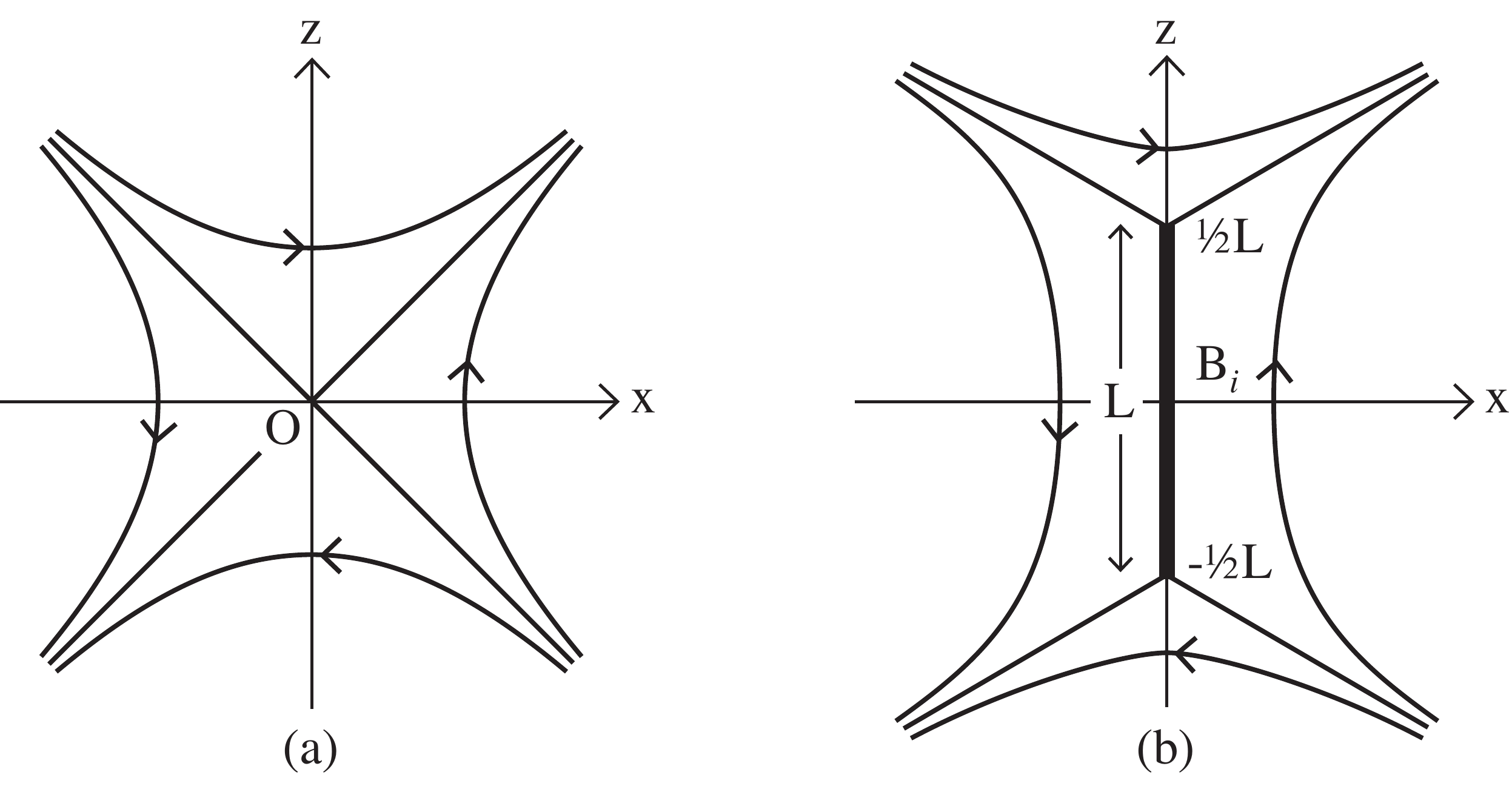}
    \caption{
    X-point magnetic field (a), which collapses into (b) a field with a current sheet of length $L$ and input magnetic field $B_i$ at $x=0+$, $z=0$.
    } 
    \label{fig1}
\end{figure}

\section{Relationship between $B_i$ and $L$ for a 2D current sheet}
\label{sec2}

Here we calculate the relationship between the input magnetic field ($B_i$) to a current sheet and its length ($L$) by firstly the traditional complex variable technique (Sect. \ref{sec2.1}) and secondly a new method that does not rely on complex variable theory (Sect. \ref{sec2.2}) and so can be generalised to 3D (Sect. \ref{sec3.2}).

\subsection{Using complex variable theory}
\label{sec2.1}

We consider a potential magnetic field of the form
\begin{equation}
B_x= kz, \ \ \ \ B_z=kx, 
\nonumber
\end{equation}
that contains an X-type neutral point at the origin, where $k$ is a constant (Fig. \ref{fig1}a). This may be written  in terms of the complex variable $Z=x+iz$ as simply
\begin{equation}
B_z+iB_x = kZ.
\nonumber
\end{equation}

Now we suppose the distant sources of the magnetic field move in such a way that the field collapses to a configuration containing a current sheet of length $L$ stretching along the $z$-axis, as shown in Fig. \ref{fig1}b. Then an elegant way of writing the field that is outside the current sheet is
\begin{equation}
B_z+iB_x = k(\textstyle{\frac{1}{4}}L^2+Z^2)^{1/2},
\label{eq3}
\end{equation}
so that the sheet is a cut in the complex plane from $Z=-\half iL$ to $Z=\half iL$.  In particular, it can be seen that the magnetic field ($B_i$) in Eq. (\ref{eq2}) at the entrance ($x=0+$, $z=0$) to the current sheet can be obtained from Eq. (\ref{eq3}) by putting $z=0$ and letting $x$ tend to zero through positive values. 
This method was first discovered by \citet*{green65} and later used by many authors, including \citet*{priest75a}, \citet*{tur76}, \citet*{somov76}, \citet*{low87,low91a}, \citet*{titov92b}.

Within a 1D current sheet with magnetic field $B_z(x)$, the electric current density is given by
\begin{equation}
\mu j=-\frac{\p B_z}{\p x}, 
\nonumber
\end{equation}
which may be integrated across a sheet of width $l$ to give a relationship between the current ($J(z)$) in the sheet at distance $z$ along it and the magnetic field $B_S(z) = B_z(\half l,z)$ at the edge of the sheet as
\begin{equation}
\mu J=\int_{-l/2}^{l/2} \mu j dx = -\int_{-l/2}^{l/2} \frac{\p B_z}{\p x} dx = -[B_z]_{-l/2}^{l/2}=-2B_z(\half l,z),
\nonumber
\end{equation}
or
\begin{equation}
\mu J(z) = -2 B_S(z).
\label{eq4}
\end{equation}

In particular, for the current sheet in Eq. (\ref{eq3}), by letting $x$ approach zero through positive values, we find
\begin{equation}
B_S(z)=k(\textstyle{\frac{1}{4}}L^2-z^2)^{1/2},
\label{eq5}
\end{equation}
and so the total current ($I$) in the sheet is given by
\begin{equation}
\mu I=\mu \int_{-L/2}^{L/2} J dz=-\int_{-L/2}^{L/2} 2k (\textstyle{\frac{1}{4}}L^2-z^2)^{1/2}dz
\nonumber
\end{equation}

or
\begin{equation}
\mu I=\textstyle{\frac{1}{4}} \pi k L^2 = -\half \pi B_i L.
\nonumber
\end{equation}

\begin{figure}[h] 
    \centering  
    \includegraphics[width=\columnwidth]{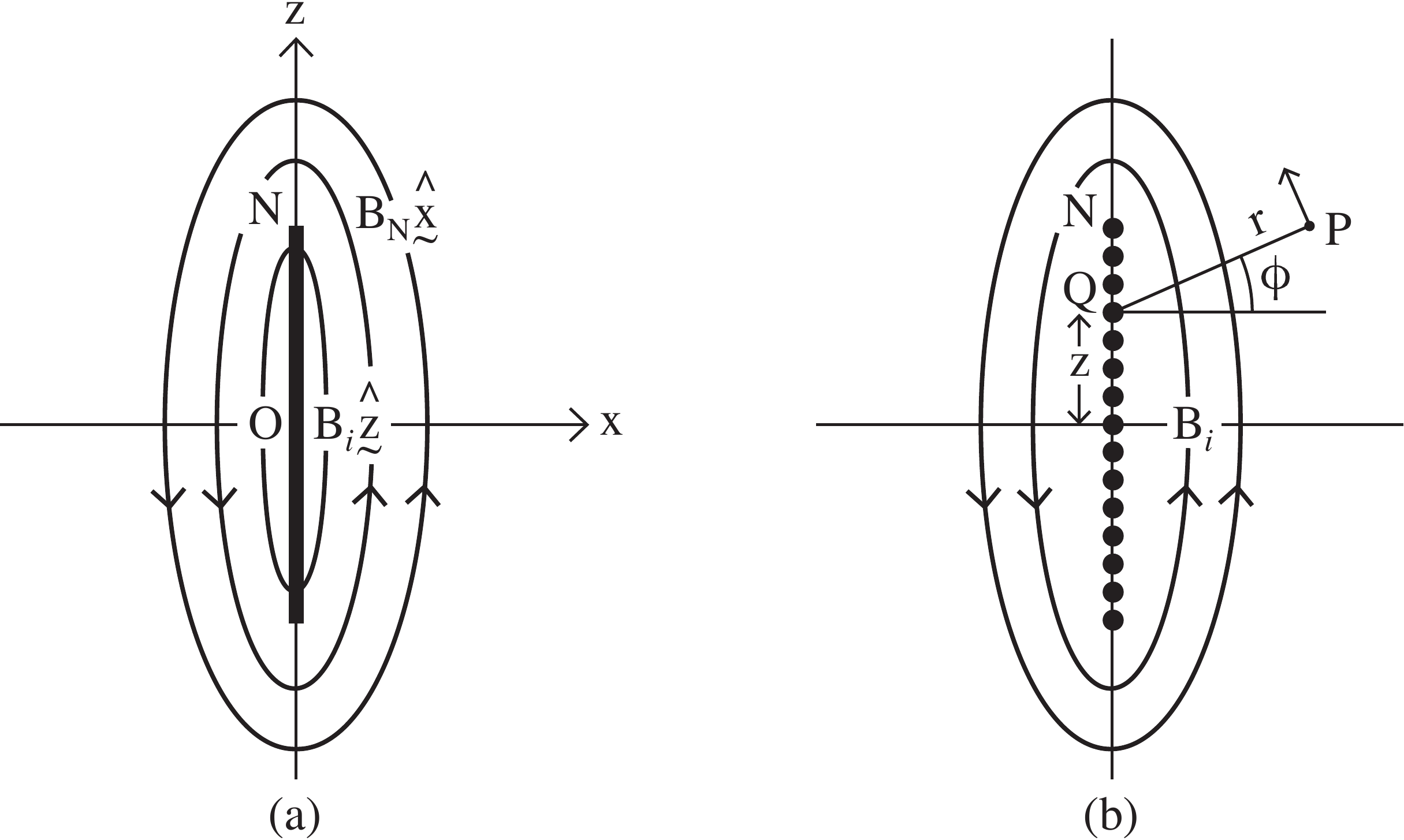}
    \caption{ Notation used for the 2D magnetic field.
    (a) Magnetic field of the current sheet alone, where the values at the points O$(0+,0)$ and N$(0,\half L)$ are denoted by $B_i \hat{\bf z}$ and $B_N \hat{\bf x}$, respectively. 
    (b) Representation of the current sheet by a set of line currents, denoted by dots, in which the magnetic field at a point P is calculated due to a line current at Q at a distance $z$ along the $z$-axis.}
    \label{fig2} 
\end{figure}
We note that the magnetic field of the current sheet alone (Fig. \ref{fig2}) is obtained by subtracting the background field from Eq. (\ref{eq3}) to give
\begin{equation}
B_z+iB_x = k(\textstyle{\frac{1}{4}}L^2+Z^2)^{1/2}-kZ,
\label{eq6}
\end{equation}
which implies that the field at ($x=0+,z=0$) is $B_z=\half kL$, while the field at the end N$(0,\half L)$ of the current sheet is $B_x=-\half kL$, namely, minus the X-point field at N, since the field of the current sheet plus background (Eq. (\ref{eq3})) vanishes at N.

\subsection{Without invoking complex variable theory}
\label{sec2.2}
We suppose the field has components
\begin{equation}
(B_x,B_z)=(kz,kx)+(b_{Sx}(x,z),b_{Sz}(x,z))
\label{eq7}
\end{equation}
due to the X-field together with the field of the current sheet.
We write the $z$-component of the field at the edge of the sheet as before, as
\begin{equation}
B_S(z)=b_{Sz}(0+,z),
\nonumber
\end{equation}
such that
\begin{equation}
B_S(0)=B_i.
\nonumber
\end{equation}
Then the current in the sheet is given as before by
\begin{equation}
\mu J(z)=-2B_S(z).
\nonumber
\end{equation}

Now we suppose the current sheet consists of an infinite set of line currents $J(z^{\prime})dz^{\prime}$ at points Q a distance $z^{\prime}$ along the current sheet, each of them producing a magnetic field, at a point P, of 
\begin{equation}
{\bm b}_S=\frac{\mu J dz^{\prime}}{2\pi r}\hat {\bm {\phi}}, 
\nonumber
\end{equation}
where  ($r,\phi,z$) are cylindrical polar coordinates measured locally relative to Q (Fig. \ref{fig2}b).

Therefore, the field at P due to the whole current sheet has an $x$-component
\begin{equation}
b_{Sx}(x,z)=-\frac{1}{\pi}\int_{-L/2}^{L/2}\frac{B_S(z^{\prime})\sin \phi}{[x^2+(z-z^{\prime})^2]^{1/2}}dz^{\prime},
\nonumber
\end{equation}
where $B_s=\half\mu J$ and $\sin \phi = (z-z^{\prime})/[x^2+(z-z^{\prime})^2]^{1/2}$.

In terms of the dimensionless variables ${\bar z}=2z/L$, ${\bar x}=2x/L$ and ${\bar B_S}=B_S/(kL)$, this becomes
\begin{equation}
{\bar b}_{Sx}({\bar x},{\bar z})=-\frac{1}{\pi}\int_{-1}^{1}\frac{{\bar B}_S({\bar z}^{\prime})({\bar z}-{\bar z}^{\prime})}{{\bar x}^2+({\bar z}-{\bar z}^{\prime})^2}d{\bar z}^{\prime}.
\label{eq8}
\end{equation}

Now the condition that our infinite set of line currents comprises a current sheet is that the tangential field vanish at its surface, meaning, that $B_x({\bar x},{\bar z}) = k{\bar z}+b_{Sx}({\bar x},{\bar z})$ vanish as ${\bar x}$ tends to zero.  In dimensionless variables this becomes
\begin{equation}
\lim_{{\bar x} \to 0}\ ({{\bar b}_{Sx}({\bar x},{\bar z})}) = -{\bar z} \ \ \ \ {\rm when} \ {\bar x}^2<1,
\nonumber
\end{equation}
which, after substituting for ${\bar b}_{Sx}$ from Eq. (\ref{eq8}), becomes
\begin{equation}
\lim_{{\bar x} \to 0}\frac{1}{\pi}\int_{-1}^{1}\frac{{\bar B}_S({\bar z}^{\prime})({\bar z}-{\bar z}^{\prime})}{{\bar x}^2+({\bar z}-{\bar z}^{\prime})^2}d{\bar z}^{\prime}=\half {\bar z}.
\label{eq9}
\end{equation}
This is an integral equation to solve for the unknown function ${\bar B}_S({\bar z}^{\prime})$.

The way we solve it is to consider $N+1$ equally spaced points ${\bar z}_0=-1,\ {\bar z}_1,\ {\bar z}_2,\ ....,\ {\bar z}_{N-1},\ {\bar z}_{N}=1$ between -1 and +1, such that ${\bar B}_S({{\bar z}_n})=B_n$ and $B_0=B_N=0$. We approximate ${\bar B}_S({\bar z}^{\prime})$  by N  linear functions in these N intervals stretching between ${\bar z}^{\prime}_n$ and ${\bar z}^{\prime}_{n+1}$, namely,
\begin{equation}
f_n({\bar z}^{\prime})=a_n{\bar z}^{\prime}+b_n = a_n({\bar z}^{\prime}-{\bar z})+c_n,
\nonumber
\end{equation}
 where $n=0,1,2,...N-1$ and $c_n=B_n+a_n{\bar z}$.  The constants that make this piecewise linear function continuous are
\begin{equation}
a_n=\frac{B_{n+1}-B_n}{{\bar z}_{n+1}-{\bar z}_n}, \ \ \  b_n=\frac{B_n{\bar z}_{n+1}-B_{n+1}{\bar z}_n}{{\bar z}_{n+1}-{\bar z}_n},
\nonumber
\end{equation}
where ${\bar z}_{n+1}-{\bar z}_n=2/N$.

\begin{figure}[h] 
    \centering  
    \includegraphics[width=\columnwidth]{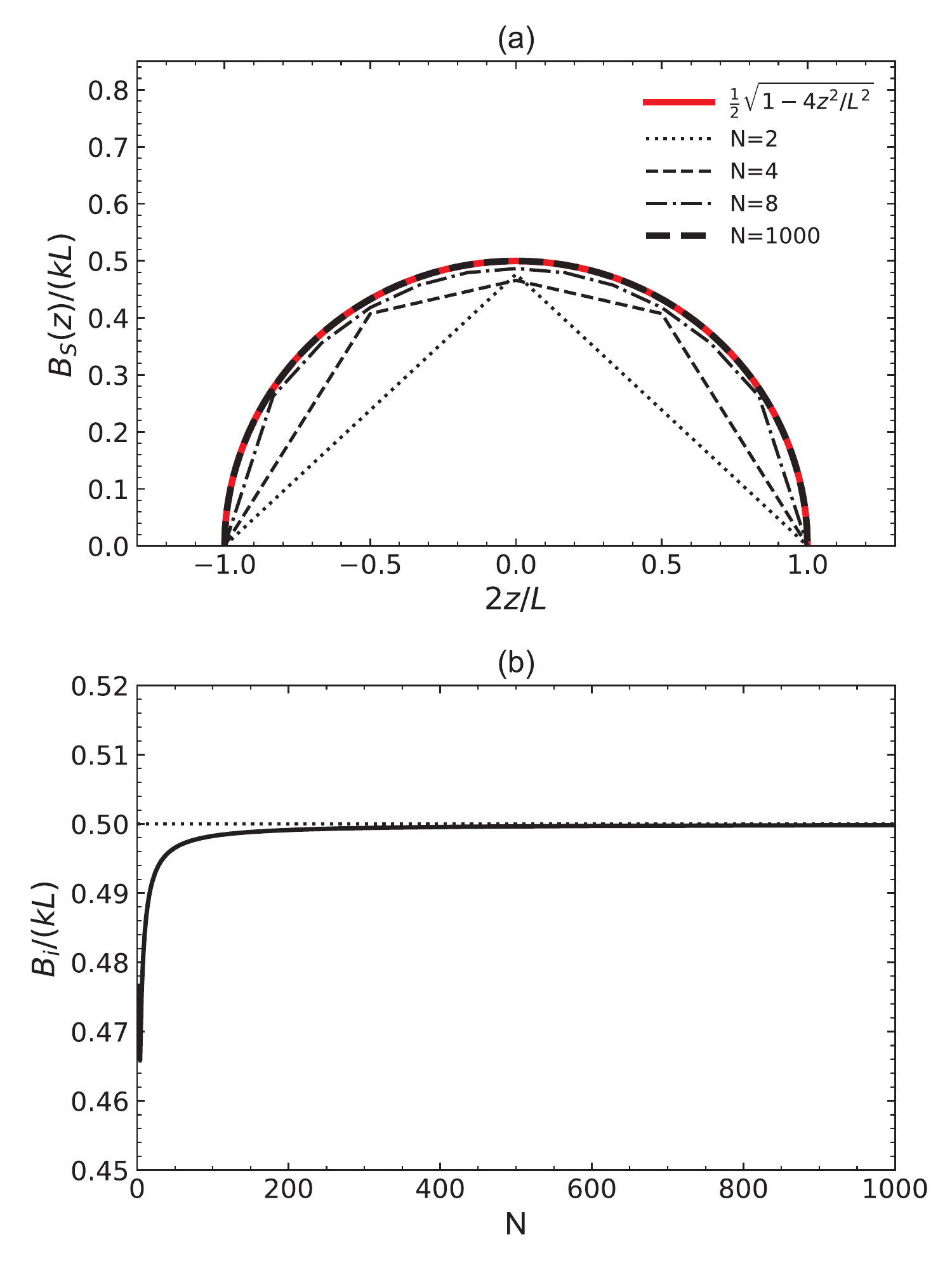}
    \caption{For a 2D current sheet (a) the dependence of the maximum current sheet field ($B_S(0)=B_i$) on $N$ and (b) the profile ($B_S(z)$) as a function of the number ($N$) of points in the sheet. }
    \label{fig3}
\end{figure}
Eq. (\ref{eq9})  is then approximated by
\begin{equation}
\lim_{{\bar x} \to 0}\frac{1}{\pi}\sum_{n=0}^{N-1}\int_{{\bar z}_n}^{{\bar z}_{n+1}}\frac{-a_n({\bar z}-{\bar z}^{\prime})^2-c_n({\bar z}-{\bar z}^{\prime})}{{\bar x}^2+({\bar z}-{\bar z}^{\prime})^2}d{\bar z}^{\prime}=\half {\bar z}.
\nonumber
\end{equation}
After some manipulation, this can be evaluated to give
\begin{multline}
\lim_{{\bar x} \to 0}\frac{1}{\pi}\sum_{n=0}^{N-1}a_n{\bar x}\arctan\left(\frac{2{\bar x}/N}{{\bar x}^2+({\bar z}_{n+1}-{\bar z})({\bar z}_n-{\bar z})}\right)
-\frac{2a_n}{N}- \\ 
-{\half c_n}\log_e\left(\frac{{\bar x}^2+({\bar z}_{n+1}-{\bar z})^2}{{\bar x}^2+({\bar z}_n-{\bar z})^2}\right)=\half {\bar z}.
\label{eq10}
\end{multline}
as shown in Fig. \ref{fig3}.
Evaluating this at N points should determine the N unknowns $B_1,\ B_2,\ ....B_N$. For the N points we pick the midpoints ${\bar z^*_m}=\half({\bar z}_m+{\bar z}_{m+1})$ of the intervals, where $B_S=\half(B_m+B_{m+1})$, $m=0,1,...N-1$ and Eq. (\ref{eq9}) becomes
\begin{multline}
\lim_{{\bar x} \to 0}\frac{1}{\pi}\sum_{n=0}^{N-1}a_n {\bar x}\arctan\left(\frac{2{\bar x}/N}{{\bar x}^2+({\bar z}_{n+1}-\half({\bar z}_m+{\bar z}_{m+1}))({\bar z}_n-\half({\bar z}_m+{\bar z}_{m+1}))}\right) \\ 
-(B_{n+1}-B_n)
-\frac{(B_n+B_{n+1})}{4}
\log_e{\left( \frac{{\bar x}^2+({\bar z}_{n+1}-\half({\bar z}_m+{\bar z}_{m+1}))^2}{{\bar x}^2+({\bar z}_n-\half({\bar z}_m+{\bar z}_{m+1}))^2}\right)}\\
= \textstyle{\frac{1}{4}}({\bar z}_m+{\bar z}_{m+1}).
\nonumber
\end{multline}
In the limit as ${\bar x} \rightarrow 0$, the first term in the summation vanishes, while the second term reduces to $B_0-B_N$, which also vanishes and so we are left with N equations for each value of $m$ of the form
\begin{multline}
-\frac{1}{\pi}\sum_{n=0}^{N-1}
(B_n+B_{n+1})
\log_e{\left| \frac{{\bar z}_{n+1}-\half({\bar z}_m+{\bar z}_{m+1})}{{\bar z}_n-\half({\bar z}_m+{\bar z}_{m+1})}\right|}
= \textstyle{\frac{1}{2}}({\bar z}_m+{\bar z}_{m+1}).
\nonumber
\end{multline}
For each value of N, the N values of $B_m= {\bar B}_S(z_m)$ are calculated by numerically solving these equations, with the result, as shown in Fig. \ref{fig3}, that the piecewise linear approximation to ${\bar B}_S({\bar z})$ tends to the function ${\bar B}_S({\bar z})=\half \sqrt{1-{\bar z}^2}=\half \sqrt{1-4z^2/L^2}$ as $N\rightarrow \infty$, as expected from the complex variable theory result, with $B_i=\half kL$.

\begin{figure}[h] 
    \centering  
    \includegraphics[width=\columnwidth]{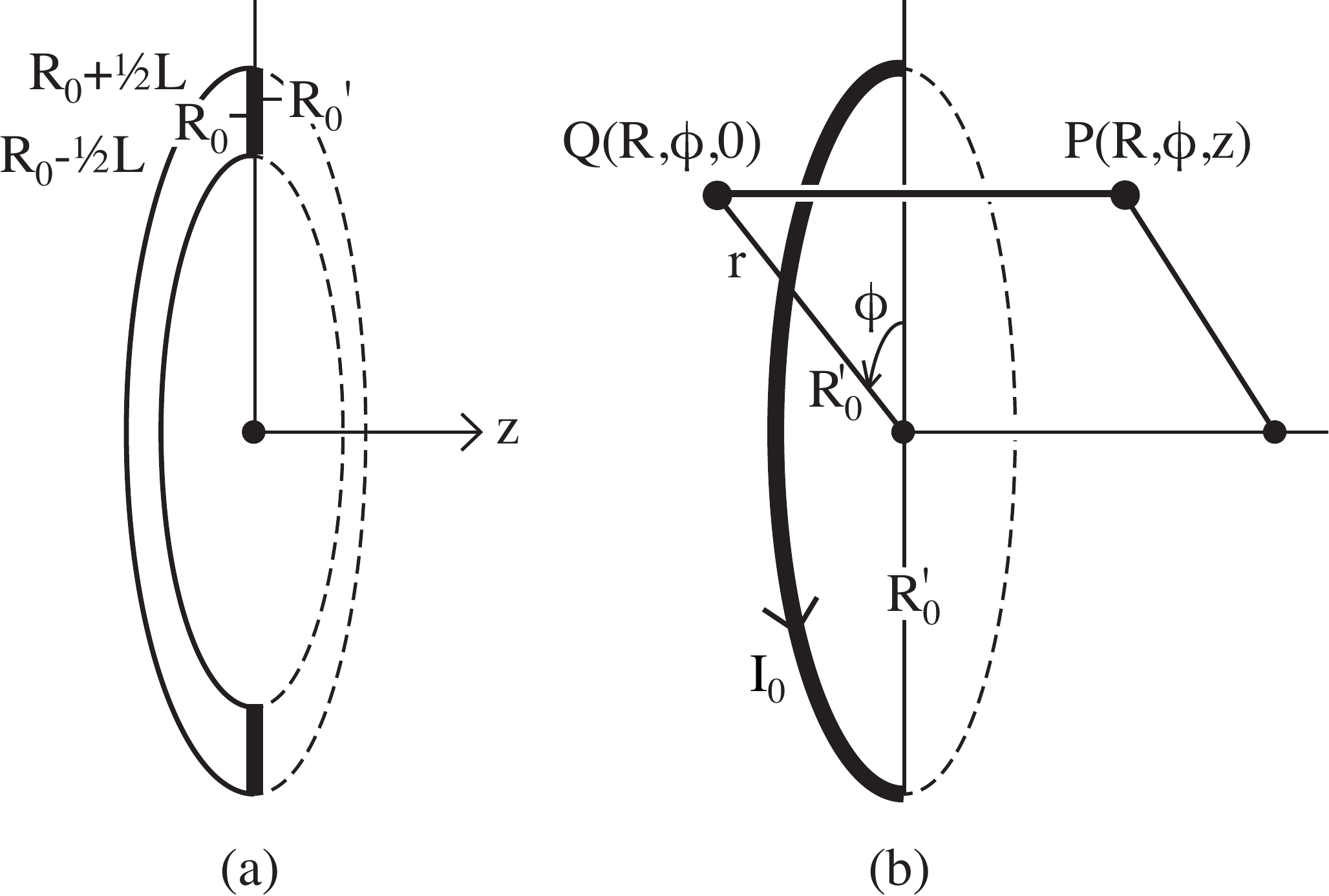}
    \caption{Notation in cylindrical polar coordinates $(R,\phi,z)$ for (a) an axisymmetric current sheet stretching from $R=R_0-\half L$ to $R=R_0+\half L$ and (b) a ring current of radius $R_0^{\prime}$ indicating the positions of a general point P$(R,\phi,z)$ and a point Q$(R,\phi,0)$ in the plane of the ring at a radius $R=R_0^{\prime}+r$. }
    \label{fig4}
\end{figure}

\section{Relationship between $B_i$ and $L$ for a 3D axisymmetric current sheet}
\label{sec3}

Complex variable theory applies only to 2D. However, the analysis of Sect. \ref{sec2.2} may be extended into 3D in order to calculate the magnetic field of a 3D axisymmetric current sheet (Fig. \ref{fig4}a), 
as first suggested by \citet*{tur77} and explored briefly by \citet*{longcope96a}. The technique we propose here is built on their ideas. 
Using cylindrical polar coordinates $(R,\phi,z)$, we write the radial and axial components of the magnetic field at $P(R,\phi,z)$ near the current sheet in the form 
\begin{equation}
(B_R,B_z)=(kz,kr)+(b_{SR}(R,z),b_{Sz}(R,z)),
\label{eq11}
\end{equation}
where the first term represents the field of a ring of X-points near $R=R_0$, while $(b_{SR},b_{Sz})$ is the field of the current sheet itself.
The tangential  (namely, $R$-) component of the field at the edge of the sheet is then, say, 
\begin{equation}
B_S(R)=b_{SR}(R,0+),
\nonumber
\end{equation}
such that at the centre of the sheet 
\begin{equation}
B_S(0)=B_i.
\nonumber
\end{equation}
Also, the integral form of Amp\`ere's law may be used to show that the current $J\hat {\bm {\phi}}$ in the sheet is related to $B_S$ by
\begin{equation}
\mu J(R)=2B_S(R).
\label{eq12}
\end{equation}

The aim is to deduce what profile of $B_S$ and therefore of current $J$ in the sheet makes the normal component ($B_z(R_0^{\prime})$) of magnetic field vanish at the current sheet so that, according to Eq. (\ref{eq11}),
\begin{equation}
kR=-\lim_{{z} \to 0} b_{Sz}(R,z) \ \ \ \  {\rm for} \ \ \ \ z=0, \ \ -\half L<R_0^{\prime}<\half L. 
 \label{eq13}
 \end{equation}
The plan is therefore to calculate the magnetic field due to a current ring (Sect. \ref{sec3.1}),  and then to sum over an infinite set of infinitesimal current rings to find the magnetic field  of the current sheet (Sect. \ref{sec3.2}).

\subsection{The magnetic vector potential for a toroidal current ring}
\label{sec3.1}

We consider a ring of current ($I_0(R_0^{\prime})$) of radius $R_0^{\prime}$ in the $z=0$ plane in cylindrical polar coordinates (Fig. \ref{fig4}b). The flux function [$F(R,\phi)$] at a point $P(R,\phi,z)$ may be calculated as follows \citep{jackson99}.  In general, the vector potential ($\bA$) is such that $\bB=\curl \bA$ and satisfies Poisson's equation 
\begin{equation}
\nabla^2\bA=-\mu \bj,
\nonumber
\end{equation} 
which has general solution 
\begin{equation}\bA=\frac{\mu}{4\pi}\int \frac{ \bj(\br^{\prime})}{|\br -\br^{\prime}|} dV^{\prime}.
\nonumber
\end{equation}  
For our current ring $\bj \ dV^{\prime}=I_0 \delta(R^{\prime}-R_0^{\prime})\delta(z^{\prime})R^{\prime} d\phi^{\prime}dz^{\prime}{\hat {\bm \phi}}$, and at $P(R,\phi,z)$ the  only component of $\bA$ is  $A_\phi(R)$, giving a flux function $F\equiv RA_\phi$ of
\begin{multline}
F(R,\phi,z)=
\frac{\mu I_0 R_0^{\prime}R}{4\pi}\int_0^{2\pi} \frac{\cos \phi^{\prime} \ d\phi^{\prime}}{{(R^2+R_0^{\prime}}^2+z^2-2R_0^{\prime}R\cos \phi^{\prime})^{1/2}},
\label{eq14} 
\end{multline}
where 
\begin{equation}
s^{\prime}=[(R-R_0^{\prime}\cos \phi^{\prime})^2+(R_0^{\prime})^2\sin^2 \phi^{\prime}+z^2]^{1/2}
\nonumber
\end{equation} 
is the distance between the points $(R,0,z)$ in the plane $\phi=0$ and $(R_0^{\prime},\phi^{\prime},0)$  on the ring.  The corresponding field components are
\begin{equation}
B_R=-\frac{1}{R}\frac{\p F}{\p z},\ \ \ \ \ \ B_z=\frac{1}{R}\frac{\p F}{\p R}.
\nonumber
\end{equation}

After some manipulation, Eq. (\ref{eq14}) may be written as
\begin{equation}
F(R,\phi,z)=\frac{\mu I_0}{2\pi k}\sqrt{R_0^{\prime}R}\ [(2-K^2)M(K)-2E(K)],
\label{eq15}
\end{equation}
in terms of $K$  defined by
\begin{equation}
K^2=\frac{4R_0^{\prime}R}{(R+R_0^{\prime})^2+z^2},
\label{eq16}
\end{equation}
and the complete elliptic integrals of the first and second kind
\begin{eqnarray}
M(K)=\int_0^{\pi/2}[1-K^2\sin^2x]^{-1/2}dx,\nonumber \\
E(K)=\int_0^{\pi/2}[1-K^2\sin^2x]^{1/2} dx. 
\nonumber
\end{eqnarray}

\begin{figure}
    \centering
    \includegraphics[width=5cm]{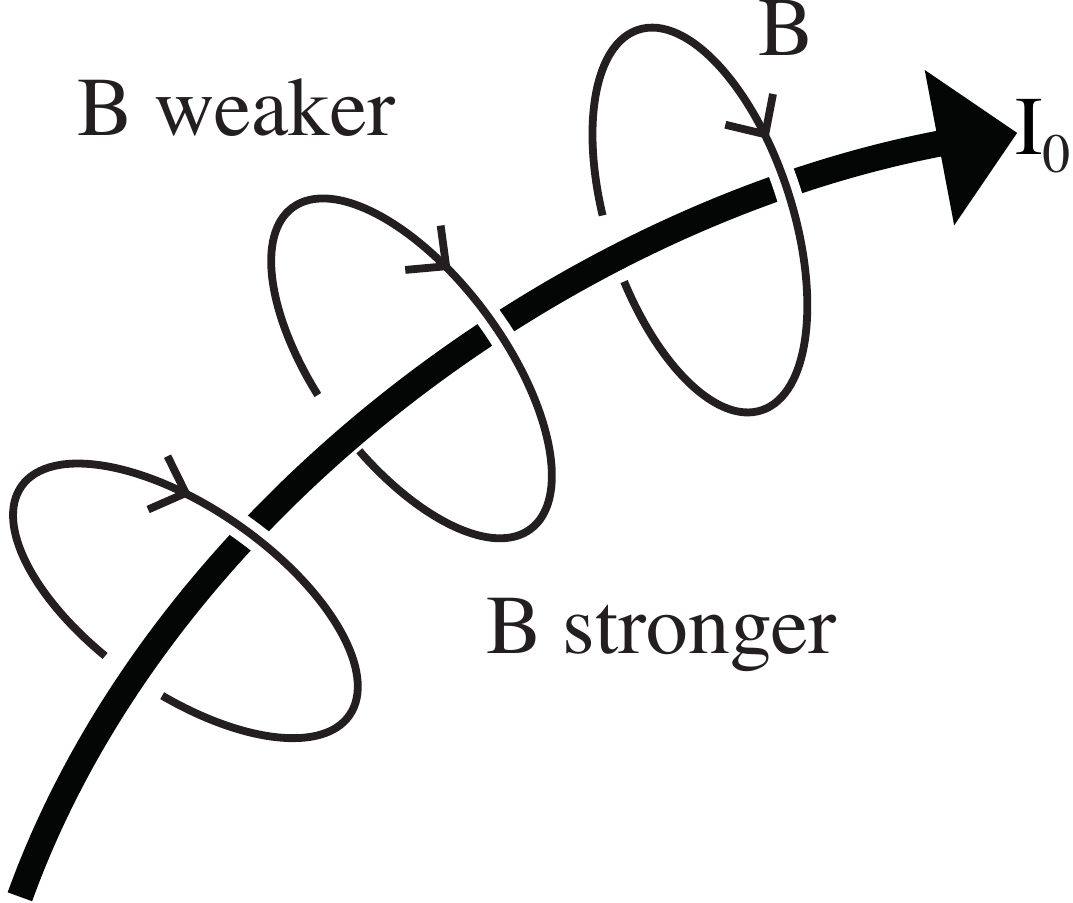}
    \caption{Magnetic field lines near a segment of a circular current loop of current $I_0$.}
    \label{fig5}
\end{figure}

The flux function at P($R,\phi,z$) near the current ring (Fig. \ref{fig4}b) can be found by 
writing $R=R_0^{\prime}+r$ and expanding Eq. (\ref{eq16})  in powers of $r/R_0^{\prime}\ll1$ and $z/R_0^{\prime}\ll1$, using
 \citet*{gradshteyn80}, to give
\begin{multline}
{F=RA}=\frac{\mu I_0}{4\pi}\\
\left\{2R_0^{\prime}\left(\log_e\frac{8R_0^{\prime}}{(z^2+r^2)^{1/2}}-2\right)+r\left(\log_e\frac{8R_0^{\prime}}{(z^2+r^2)^{1/2}}-1\right)\right\}  + . . . .
\label{eq17}
\end{multline}
The corresponding magnetic field ($B_z=(1/R)\p F/\p R$)  close to the ring  becomes, to lowest order in $(r^2+z^2)^{1/2}/R_0^{\prime}$,
\begin{equation}
B_z \approx -\frac{\mu I_0}{4\pi}\left\{\frac{2r}{r^2+z^2}+\frac{1}{R_0^{\prime}}\log_e\frac{(z^2+r^2)^{1/2}}{8R_0^{\prime}}\right\},
\label{eq18}
\end{equation}
where $r=R-R_0^{\prime}$.
The first term  is simply the field of a straight current, and the second term gives the correction due to the curvature of the current ring. This correction lowers the magnitude of the field on the outside of the ring and increases it on the inside of the ring, as expected, since the field lines are further apart on the outer edge of the ring, as illustrated in Fig. \ref{fig5}.

\begin{figure}[h] 
    \centering  
    \includegraphics[width=\columnwidth]{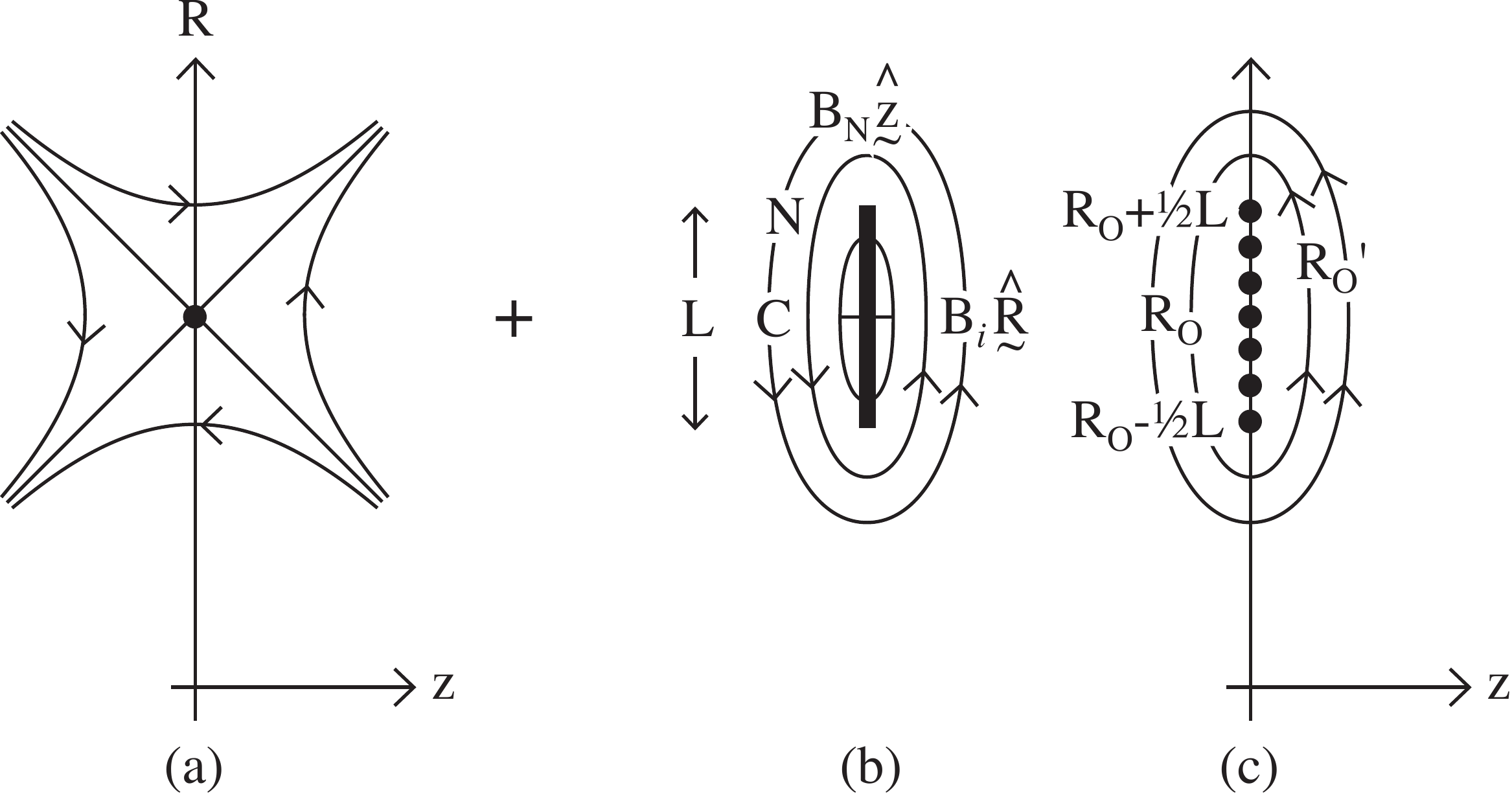}
    \caption{Notation for the magnetic field near a current sheet in the $Rz$-plane due to the sum of  (a) a ring of nulls at radius $R_0^{\prime}$ and (b) a current sheet of length $L$. The sheet may be replaced by (c) a set of current rings  of radius $R_0^{\prime}$ between $R=R_0-\half L$ and $R=R_0+\half L$. Just to the right of the centre C of the sheet, the magnetic field is $B_i \bf \hat{R}$, while at the outer edge N of the current sheet the field is $B_N\bf \hat{z}$.}
    \label{fig6}
\end{figure}

\subsection{The magnetic field of a current sheet}
\label{sec3.2}

We consider a current sheet of length $L$ in the $R$-direction and centred at a radius $R_0$ with $L\ll R_0$, so that the current sheet is short compared with its mean distance $R_0$ from the $z$-axis, as shown in Fig. \ref{fig6}. It is made up of  an infinite set of infinitesimal current rings that are located at radius ${R_0}^{\prime}$, say, with currents $J(R_0^{\prime})\delta(R-R_0^{\prime}) dR_0^{\prime} \hat {\bm {\phi}}$, where $R_0^{\prime}=R_0+r_0^{\prime}$ ranges between $R_0-\half L$ and $R_0+\half L$, as indicated in Fig. \ref{fig6}c, where $r_0^{\prime}\ll R_0$. Each ring has a flux function of the form of Eq. (\ref{eq17}) and a $z$-component of magnetic field of the same form as Eq. (\ref{eq18}).   

The magnetic field of the current sheet may now built up an integral of infinitesimal current rings $J(R_0^{\prime})dz$ at $R_0^{\prime}=R_0+r_0^{\prime}$, 
where $\mu J=2B_S$ and each current ring gives rise to a magnetic field of the form of Eq. (\ref{eq18}) with the distance between the current ring and P$(R,\phi,z)$ being obtained by replacing $r$ in Eq. (\ref{eq18}) by $(r-r_0^{\prime})$. The resulting $z$-component of magnetic field is
\begin{multline}
b_{Sz}(R,z) \approx -\frac{1}{\pi}\int_{-L/2}^{L/2}\half B_S(r_0^{\prime})\\
\left\{\frac{2(r-r_0^{\prime})}{z^2+(r_0^{\prime}-r)^2}+\frac{1}{R_0}\log_e\frac{[z^2+(r_0^{\prime}-r)^2)]^{1/2}}{8R_0}\right\}dr_0^{\prime}.
\label{eq19}
\end{multline}

Then, after using this expression, the condition (Eq. (\ref{eq13})) that the normal magnetic component vanish at the current sheet becomes
\begin{multline}
kR=\frac{1}{\pi}\lim_{{z} \to 0}  \int_{-L/2}^{L/2}\half B_S(r_0^{\prime})\\
\left\{\frac{2(r-r_0^{\prime})}{z^2+(r_0^{\prime}-r)^2}+\frac{1}{R_0}\log_e\frac{[z^2+(r_0^{\prime}-r)^2)]^{1/2}}{8R_0}\right\}dr_0^{\prime},
\nonumber
\end{multline}
or, in terms of dimensionless variables
\begin{equation}
{\bar z}=2z/L, \ \  {\bar r_0^{\prime}}=2r_0^{\prime}/L,\ \  {\bar r}=2r/L,\ \  {\bar B_S}=B_S/(k L),
\nonumber
\end{equation}
\begin{multline}
{\bar r}=\frac{1}{\pi}\lim_{{\bar z} \to 0}\int_{-1}^{1} {\bar B_S}({\bar r_0^{\prime}}) \\
\left \{\frac{2( {\bar r}-{\bar r_0^{\prime}})}{{\bar z}^2+({\bar r_0^{\prime}}-{\bar r})^2}+\epsilon \log_e\left(\frac{\epsilon}{8}[{\bar z}^2+({\bar r_0^{\prime}}-{\bar r})^2]^{1/2}\right) \right \}d{\bar r_0^{\prime}},
\label{eq20}
\end{multline}
where $\epsilon = L/(2R_0)\ll 1$. After taking the limit as ${\bar z}$ tends to zero, this reduces to
\begin{equation}
{\bar r}=\frac{1}{\pi}\int_{-1}^{1} {\bar B_S}({\bar r_0^{\prime}}) 
\left \{\frac{2}{( {\bar r}-{\bar r_0^{\prime}})}+\epsilon \log_e\frac{\epsilon}{8}
+\epsilon \log_e | {\bar r_0^{\prime}}-{\bar r}| \right \}d{\bar r_0^{\prime}}.
\label{eq21}
\end{equation}
Using the fact that the last term is important only where $| {\bar r_0^{\prime}}-{\bar r}| \lesssim \epsilon/8$, so that $ {\bar B_S}({\bar r_0^{\prime}})\approx  {\bar B_S}({\bar r_0})$, we may evaluate its integral to give
\begin{equation}
{\bar r}=\frac{1}{\pi}\int_{-1}^{1} {\bar B_S}({\bar r_0^{\prime}}) 
\left \{\frac{2}{( {\bar r}-{\bar r_0^{\prime}})}+\epsilon \log_e\frac{\epsilon}{8}\right \}d{\bar r_0^{\prime}}
+\frac{1}{\pi} {\bar B_S}({\bar r})\frac{\epsilon}{4}\log_e\frac{\epsilon}{8}.
\label{eq22}
\end{equation}

This is an integral equation for the unknown function ${\bar B_S}({{\bar r_0^{\prime}}})=B_S(r_0^{\prime})/( kL)$, which may be solved, as in the 2D case, by approximating the function by a piecewise linear function of $N$ equally spaced straight lines (${\bar B_S}(s)=a_ns+b_n$) with $N$ constants $B_n$, and evaluating it at $N-1$ values $S_i=\half (s_m+s_{m+1})$, where $i=0,1,2,....(N-1)$. 

\begin{figure}
    \centering
    \includegraphics[width=\columnwidth]{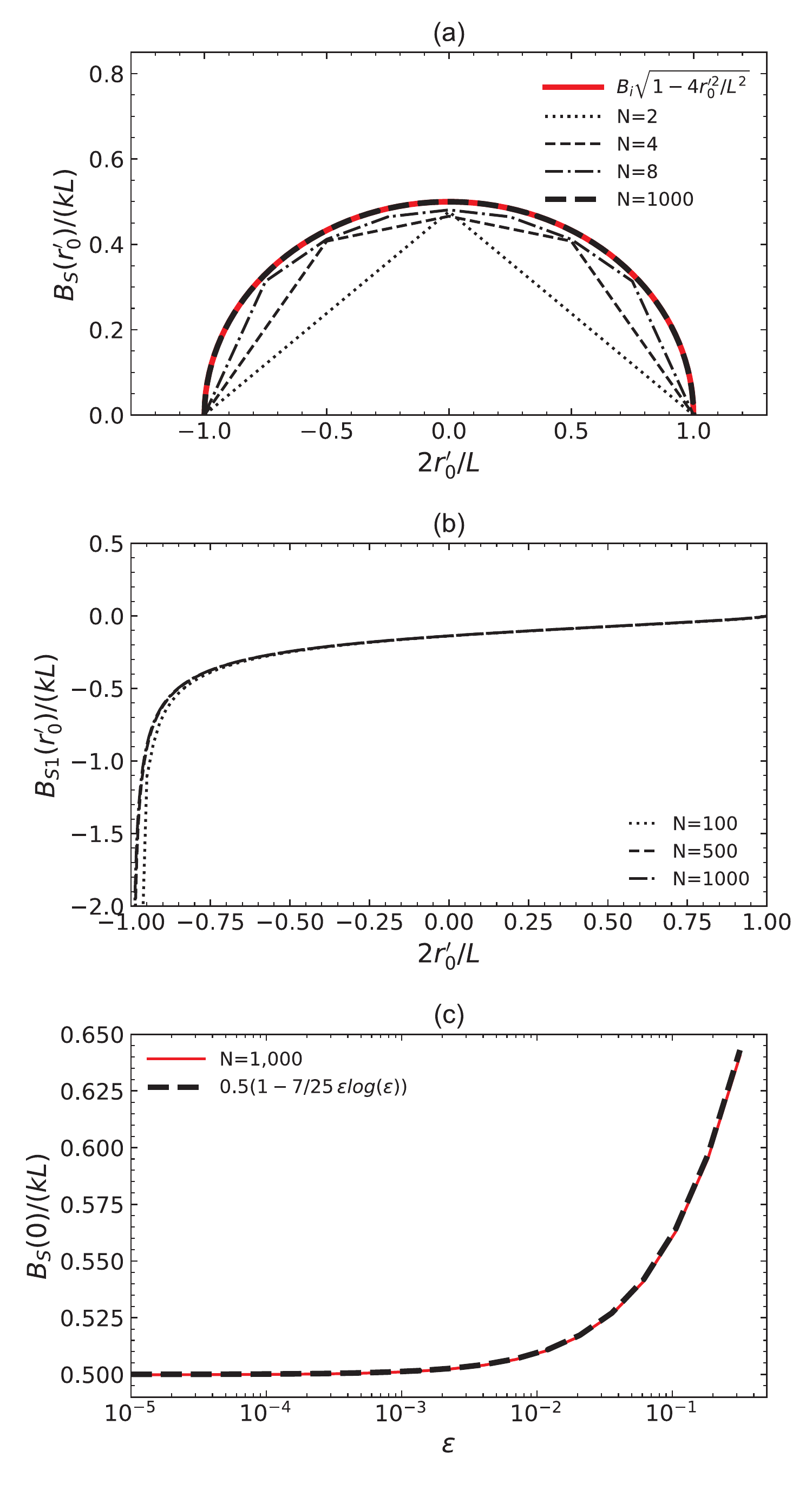}
    \caption{For a  current sheet in 3D, the profiles as functions of radius $R_0^{\prime}$ and the number $N$ of points in the sheet of (a)  the zeroth order field $B_S(R_0^{\prime})$  and (b)  the functional form ($B_{S1}(R_0^{\prime})$ of the toroidal correction. (c) The inflow magnetic field to the current sheet, namely, $B_i= B_S(0)$ as a function of $\epsilon=L/(2R_0)$ for large $N$ and its analytical approximation (dashed).}
    \label{fig7}
\end{figure}

However, Eq. (\ref{eq22}) implies that the natural expansion parameter is $\epsilon \log_e\epsilon/8\ll 1$ rather than $\epsilon$, and so we may write
 \be
 {\bar B_S}\approx {\bar B_{S0}}+\epsilon\log_e(\epsilon/8) {\bar B_{S1}},
 \nonumber
 \ee
where ${\bar B_{S0}}$ is the straight-field contribution and  ${\bar B_{S1}}$ is the toroidal correction, where $\epsilon = L/(2 R_0)\ll 1$. The zeroth and first order parts of Eq. (\ref{eq22}) become
\begin{equation}
{\bar r}=\int_{-1}^{1}  \frac{2{\bar B_S}({\bar r_0^{\prime}})}{| {\bar r_0^{\prime}}-{\bar r}|}d{\bar r_0^{\prime}},
\label{eq23}
\end{equation}
which is the same form as Eq. (\ref{eq9}), and
\begin{equation}
0=\int_{-1}^{1}  \left \{\frac{2{\bar B_{S1}}({\bar r_0^{\prime}})}{| {\bar r_0^{\prime}}-{\bar r}|}+ {\bar B_{S0}}({\bar r_0^{\prime}})\right \}d{\bar r_0^{\prime}}+\textstyle{\frac{1}{4}}B_{S0}({\bar r}),
\label{eq24}
\end{equation}
which determines ${\bar B_{S1}}({\bar r_0^{\prime}})$.  The solutions for ${\bar B_{S0}}$ and ${\bar B_{S1}}$ are shown in Fig. \ref{fig7}a,b.

The main aim of this section is to determine the inflow field ($B_i$) to the current sheet. In the 2D case, it is just $\half kL$. For the toroidal current sheet, it becomes
\begin{equation}
B_i =B_S(0)= kL{\bar B_{S0}}(0)\{1+ \epsilon\log_e\epsilon{\bar B_{S1}}(0)/{\bar B_{S0}}(0)\},
\nonumber
\end{equation}
where ${\bar B_{S0}}=\half$ and ${\bar B_{S1}}(0)/{\bar B_{S0}}(0)=-0.2757\approx -7/25$, so that
\begin{equation}
B_i = \half kL(1- 0.2757\ \epsilon\log_e\epsilon).
\label{eq25}
\end{equation}
The resulting variation of $B_i$ with $\epsilon=L/(2R_0)$ is plotted in Fig. \ref{fig7}c, which shows how $B_i$ increases as the radius $R_0$ decreases and indicates the excellence of the 7/25 approximation.

\begin{figure}[h]
    \centering \includegraphics[width=\columnwidth]{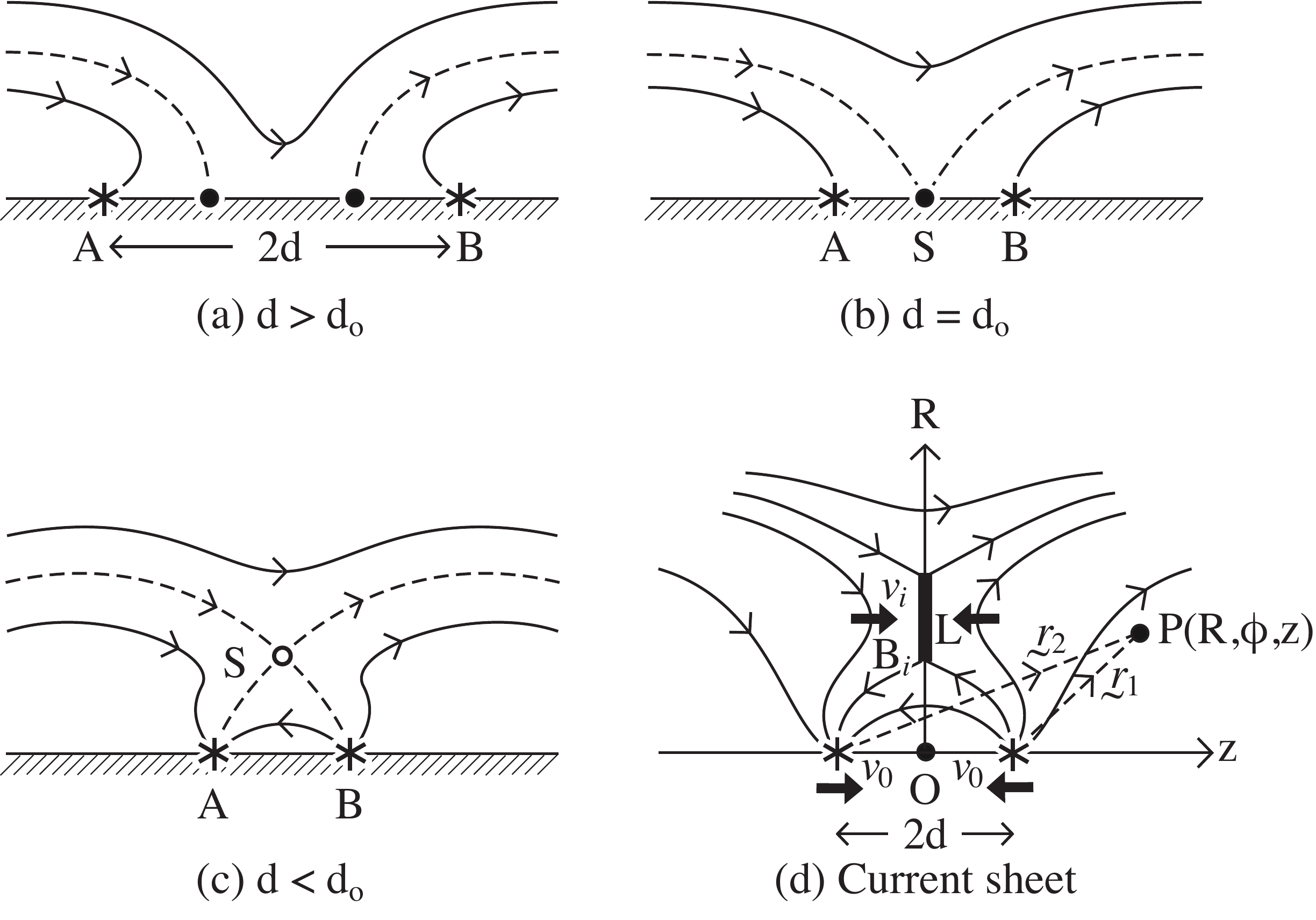}
    \caption{
    Phase 1 of the cancellation process. 
    (a) Two photospheric magnetic sources of flux $\pm F$, situated on the $z$-axis a distance $2d$ apart in an overlying uniform horizontal field $B_0{\bf \hat z}$ approach one another at speed $\pm v_0$. 
    (b) When $d=d_0$, a separator S is formed. 
    (c) Reconnection is driven at the separator S which rises in the atmosphere. 
    (d) Energy is converted at a  current sheet of length $L$, where plasma flows in at speed $v_i$ carrying magnetic field $B_i$.
 } 
    \label{fig8}
\end{figure}

\section{Reconnection driven by the approach of two magnetic fragments in a uniform horizontal field}
\label{sec4}

Here we develop, in several ways, the theory for reconnection driven by the approach and cancellation of two photospheric magnetic fragments that was proposed in \citet*{priest18}. The fragments have  equal but opposite magnetic flux ($\pm F$) and are situated in an overlying uniform horizontal magnetic field  of strength $B_0$. They are separated by a distance $2d$ and approach each other at speeds  $\pm v_{0}$ (Fig. \ref{fig8}a). The theory so far has concerned  `Phase 1' of heating and jet acceleration, during which a separator forms in the photosphere at a critical separation,
 \begin{equation}
 d=d_0 \equiv \left(\frac{F}{\pi B_0}\right)^{1/2},
 \label{eq4.1}
 \end{equation}
 called the interaction distance \citep{longcope98}  (Fig. \ref{fig8}b). The separator is located at a height $R=R_S$ (Fig. \ref{fig8}c) which increases to a maximum value of $0.6 d_0$ and then moves back downwards, reaching the photosphere as $d\rightarrow 0$. During the rise and fall of the separator, separator reconnection is driven at a current sheet of length $L$, where the input flow speed and magnetic field  to the current sheet are  $v_i$ and $B_i$, respectively (Fig. \ref{fig8}d).  The theory estimates the values of $L$, $v_i$, and $B_i$ in terms of the imposed parameters $v_0$, $B_0$ and $F$, and shows that the heating is likely to be sufficient to heat the chromosphere and corona by a so-called `cancellation nanoflare mechanism'.

The two ways we extend the theory are: using the above analysis for a 3D current sheet during the Phase 1 (Sect. \ref{sec4.1}) rather than a 2D one; and considering briefly the nature of the heating during a new Phase 2, namely, the `cancellation phase' during which the polarities are very close to each other (Sect. \ref{sec4.2}), and the two photospheric fragments actually cancel with one another.  For simplicity, we  formulate the analysis in terms of cylindrical polar rather than rectangular Cartesian coordinates. 

We note that another possibility has been suggested by \cite{low91a}, namely, that of a `Phase 0' such that, after the separator appears in the solar surface in Fig. \ref{fig8}b, a current sheet grows upwards from the solar surface rather than being localised around a separator located above the photosphere. We shall analyse this possibility in future and compare the energy release with the case we are studying here. If the driving does not switch on and off, so that the current sheet dissipates and then reforms at a different height, or if reconnection is slow enough that the current sheet does not go unstable to tearing, it is possible that such a Phase 0 exists for some time.

\subsection{Phase 1 of cancellation}
\label{sec4.1}

The  magnetic field above the photosphere ($R>0$) may be written in terms of cylindrical polar coordinates $(R,\phi,z)$, with the $z$-axis being horizontal and situated in the photosphere, joining the two magnetic fragments located at $(0,0,\pm d)$ (Fig. \ref{fig8}a)
\begin{equation}
{\bf B}=\frac{F\ \bf{\hat{r}_{1}}}{2\pi r_{1}^{2}}-\frac{F\ \bf{\hat{r}_{2}}}{2\pi r_{2}^{2}}+B_0{\bf{\hat{z}}},
\label{eq4.2}
\end{equation}
where
\begin{equation}
{\bf r}_{1}=(z-d){\bf \hat{z}}+R\ {\bf \hat{R}}, \ \ \ \ 
{\bf r}_{2}=(z+d){\bf \hat{z}}+R\ {\bf \hat{R}} \nonumber
\nonumber
\end{equation}
are the vector distances from the two sources to a point P($R,\phi,z$).

We consider what happens when the distance $2d$ between the two sources 
decreases from a large value. When the sources are too far apart, such that $d>d_0$, 
two separatrix surfaces completely surround the 
flux that enters one source and leaves the other, so that no magnetic field lines link 
one source to another.  On the other hand, when $d=d_0$ a 
separator bifurcation occurs in which these two separatrices touch at a separator field line (S) that lies in the photospheric plane ($R=0$), as described in \citet*{priest18}.  
However, when $d<d_0$ the separator 
rises above $z=0$  and a new domain is 
created bounding magnetic flux that passes under S
and links the two sources  (Fig. \ref{fig8}c). As $d$ decreases from $d_0$ to 0, we have Phase 1, during which reconnection is driven at the separator that rises to a maximum and then falls to the photosphere. Finally, when
$d=0$, Phase 1 is over, Phase 2 begins when the actual cancellation of the photospheric fragments begins (Sect. \ref{sec4.2}).

In the case of magnetic fragments of equal magnitude that we are considering here, the magnetic field is axisymmetric about the $z$-axis 
and so there is a ring of null points at distance $R_{S}$ from the 
origin in every plane through the $z$-axis. 

Along the $R$-axis, ${B}_{R}=0$ and 
\begin{equation}
 \frac{{B}_{z}}{B_0} = 
 -\frac{d\ d_0^2}{(d^2+R^2)^{3/2}}+1.
 \label{eq4.3}
\end{equation}
The location (${R}={R}_{S}$) of the separator where $B_z$ 
vanishes is therefore given  by 
\begin{equation}
{R}_{S}^{2}={d}^{2/3}{d_0}^{4/3}-{d}^{2}.   
 \label{eq4.4}
 \end{equation}
When
${d}=d_0$, the separator is located at the origin, and, as 
${d}$ decreases, it rises along the $R$-axis to a maximum height, and  thereafter it
falls back to the origin as ${d}\rightarrow 0$. The maximum height varies with $B_0$ and $F$, but is typically about  $0.6 d_0$, and so it lies in the chromosphere, transition region or corona depending on the sizes of $F$ and $B_0$ \citep*{priest18}.

When analysing flux cancellation, the natural parameters, for each value of the 
source separation ($2d$), are the critical source 
half-separation distance ($d_{0}$), the flux source speed ($v_{0} 
\equiv {\dot d}\equiv dd/dt$) and the overlying field strength ($B_{0}$). 
On the other hand, the parameters that determine the rate of release of energy at a reconnecting current sheet (Fig. \ref{fig8}c) are the inflow speed ($v_{i}$) and magnetic field ($B_{i}$) to the current sheet and the sheet length ($L$).
We now therefore proceed to calculate them as functions of $d_0$, $v_0$ and $B_0$.

Firstly, to find $B_i$ calculate  the potential field near the separator, which can be shown from Eq. (\ref{eq4.3}) to have the form $B_{z}=kR$ to lowest order,
where
\begin{equation}
k=\frac{3[1-(d/d_0)^{4/3}]^{1/2}}{(d/d_{0})^{1/3}}\frac{B_0}{d_0}.
\label{eq4.5}
\end{equation}
When a current sheet forms, the magnetic field at the inflow to
the sheet then becomes, after substituting the above value of $k$
into Eq. (\ref{eq25}), 
\begin{equation}
\frac{B_i}{B_0}=\frac{3[1-(d/d_0)^{4/3}]^{1/2}}{2(d/d_0)^{1/3}}\frac{L}{d_0}
(1- 0.2757\ \epsilon\log_e\epsilon),
\label{eq4.6} 
\end{equation}
where $\epsilon = L/(2R_0)\ll 1$.

Next, we calculate $v_{i}$ from the rate of 
change (${\dot \psi} \equiv d\psi /dt$) of magnetic flux through the 
semicircle of radius $R_{S}$ out of the plane of Fig. \ref{fig8}c.
This rate of change of flux becomes, after using  $\bf E+\bf v \times \bf B=\bf 0$ and Faraday's Law, 
 \begin{equation}
    \frac{d\psi}{dt} = -\pi R_{S}E  =  \pi R_{S}v_{i}B_{i}.
\label{eq4.7}
  \end{equation}
However, $\psi$ may be calculated from the magnetic flux below 
$z_{S}$ through the semicircle, namely,
\begin{equation}
\psi= \int_{0}^{R_{S}}\pi R\ B_{z}\ dR=F\left[\frac{3}{2}\left(\frac{d}{d_0}\right)^{2/3}-\frac{1}{2}\left(\frac{d}{d_0}\right)^{2}-1\right],
    \nonumber
\end{equation}
which vanishes when $d=d_{0}$ and increases monotonically to a value of $F$ 
as the separation ($2d$) between the sources approaches zero. The rate of change of the flux then becomes
\begin{equation}
\frac{d \psi}{dt}= \frac{v_0F}{d_0}\left[\left(\frac{d}{d_0}\right)^{-1/3}-\frac{d}{d_0}\right].
\label{eq4.8}
\end{equation}
After substituting into Eq. (\ref{eq4.7}) together with the values of $R_S$ and $B_i$ from Eqs. (\ref{eq4.4}) and (\ref{eq4.6}), the required expression for $v_i$ becomes
\begin{equation}
v_i=\frac{2v_0d_0}{3L}\left(\frac{d_0}{d}\right)^{1/3} 
\{1- 0.2757\ \epsilon\log_e\epsilon\}^{-1}.
  \label{eq4.9}
\end{equation}

\begin{figure}[h]
    \centering 
    \includegraphics[width=\columnwidth]{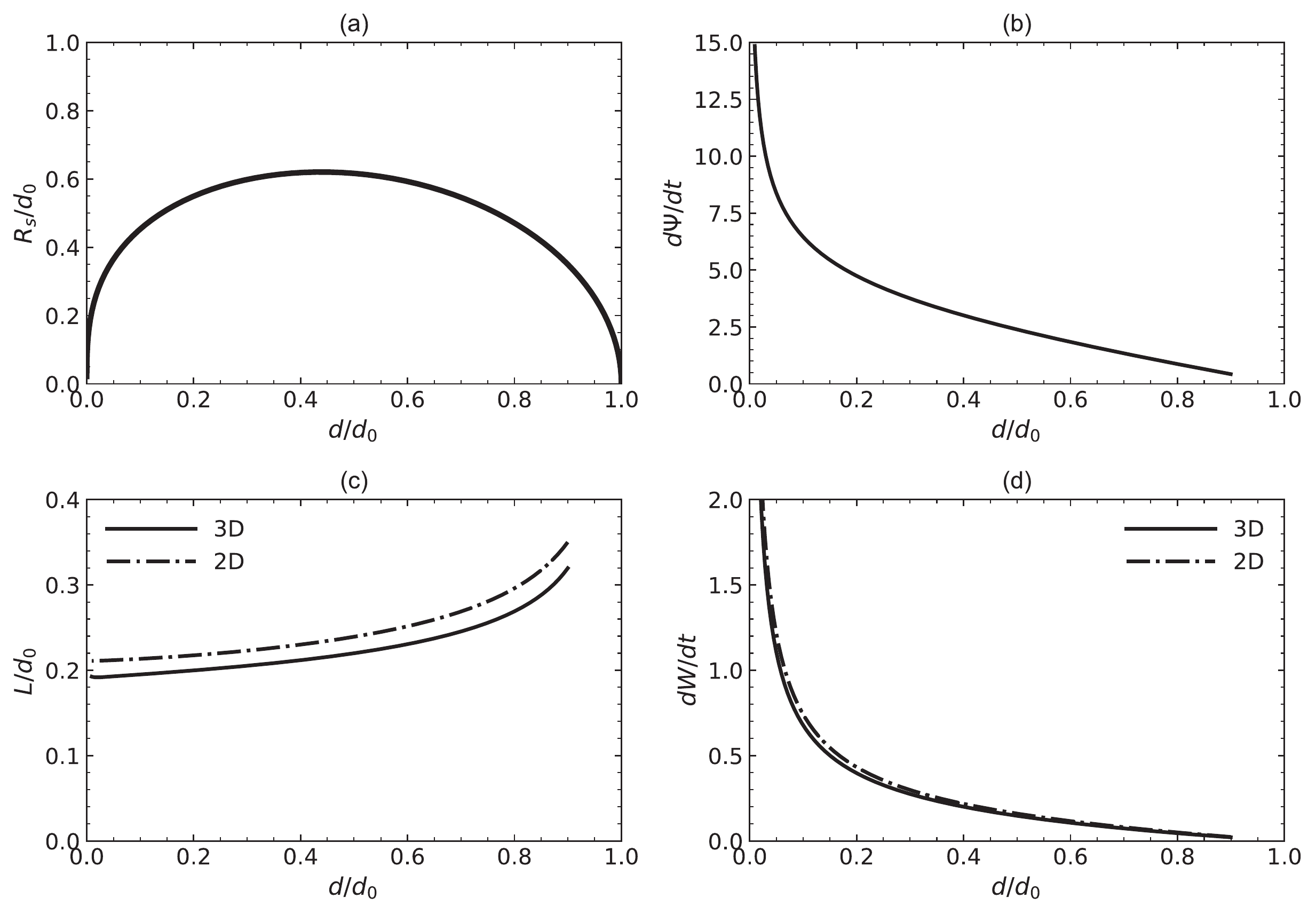}
    \caption{Phase 1 properties for a 3D current sheet as functions of the half-separation ($d$) of the two magnetic sources in units of the interaction distance ($d_0$): 
    (a) the height ($R_S$) of the separator, 
    (b) the rate of change $(d\psi/dt$) of magnetic flux below the separator, 
    (c) the current sheet length ($L$), and 
    (d) the energy conversion rate ($dW/dt$) in units of $W_0/t_0=v_0B_0^2d_0^2/\mu$. 
    Dash-dot curves show the results for a 2D sheet  \citep{priest18}.}
    \label{fig9}
\end{figure}
Then,  the rate of conversion of inflowing magnetic energy into
heat can be written, following \citet*{priest18} as
\begin{equation}
    \frac{dW}{dt}=0.8\frac{v_{i}B_{i}^{2}}{\mu}L\pi R_{S},
\label{eq4.10}
\end{equation}
where $L$ is determined  by the condition for fast reconnection that the inflow speed  acquire any value up to a maximum of
\begin{equation}
v_i = \alpha v_{Ai},
\label{eq4.20}
\end{equation} 
where $\alpha$ is 
likely to be a non-trivial function of the external parameters \citep[for example,][]{priest14a} but, as discussed in \cite{syntelis19} is
typically 0.1.
Then, after writing $v_{Ai}=v_{A0}B_{i}/B_{0}$, where $v_{A0}=B_0/\sqrt{\mu \rho_i}$,  
and substituting for  $v_i$ from Eq. (\ref{eq4.9}) and $B_i$ from Eq. (\ref{eq4.6}),  
we obtain
\begin{equation}
\frac{L^2}{d^2_0}=\frac{4M_{A0}}{9\alpha} \frac{1}{ [1 - (d/d_0)^{4/3}]^{1/2} } 
\{1- 0.2757\ \epsilon\log_e\epsilon\}^{-2},
\label{eq4.21}
\end{equation}
where $M_{A0}=v_0/v_{A0}$.
Thus, by substituting for $v_{i}$, $B_{i}$, $L$ and $R_S$ from  Eqns.(\ref{eq4.9}), (\ref{eq4.6}), (\ref{eq4.21}), and (\ref{eq4.4}),
the energy conversion rate becomes
finally
\begin{eqnarray}
    \frac{dW}{dt}=\frac{1.6\pi v_{0} B_{0}^{2}d_{0}^{2}M_{A0}[1-(d/d_0)^{4/3}]}{3\mu\alpha(d/d_0)^{2/3}(1- 0.2757\ \epsilon\log_e\epsilon)}.
\label{eq4.22}
\end{eqnarray}

The variations of $L/d_0$ and $dW/dt$ with $d/d_0$ are shown in Fig. \ref{fig9} for both the 2D and 3D cases. The curves are cut off near $d=0$ and $d=d_0$, where the analysis fails since it implies unphysically that $d>R_S$. 
The 3D treatment of the current sheet produces a correction of $9\%$ in the total energy release.
\begin{figure}[h]
    \centering \includegraphics[width=\columnwidth]{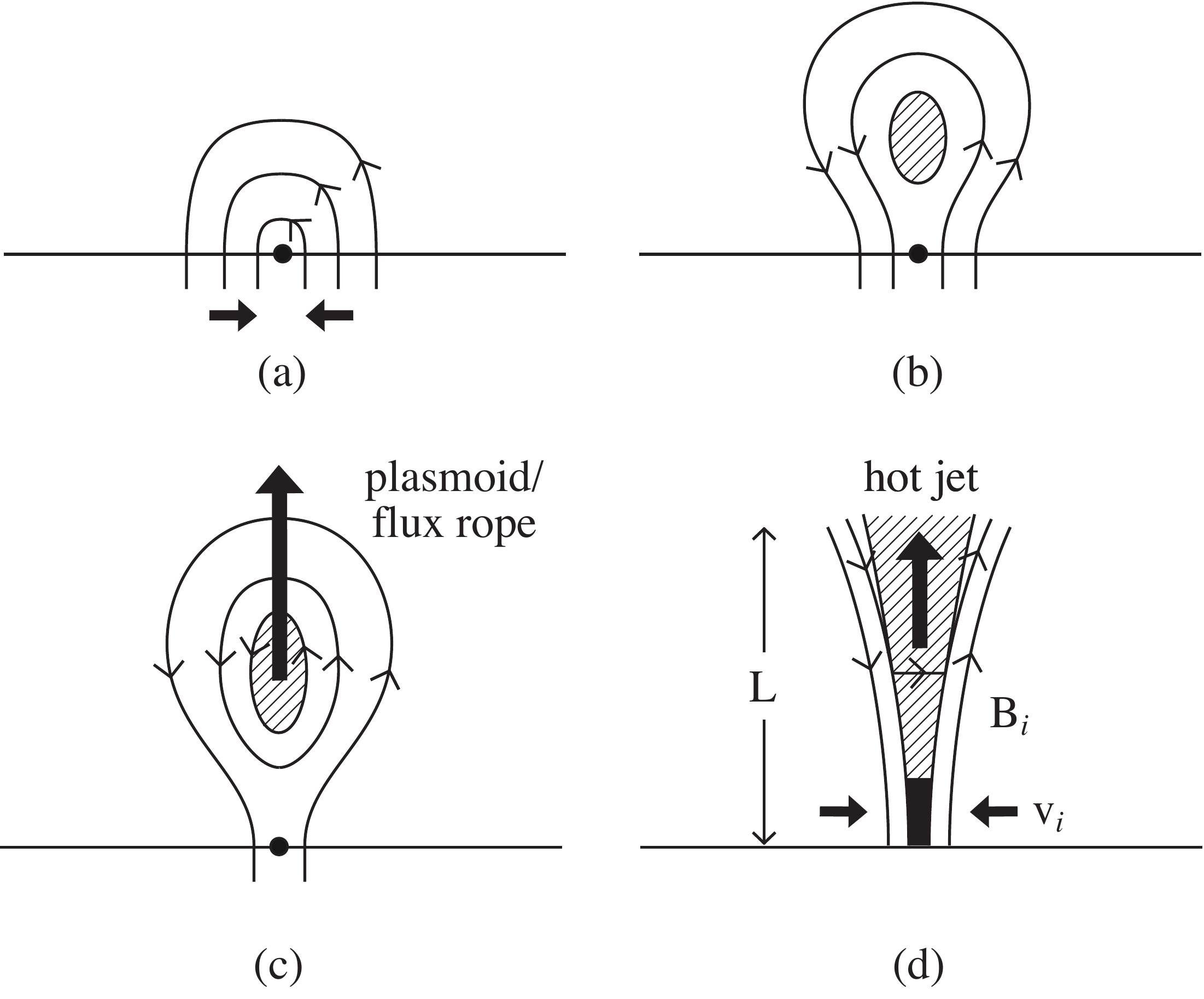}
    \caption{
    Phase 2 of cancellation.  
    (a)  Fluxes come into contact in the photosphere and reconnection is driven there, creating (b) a flux rope (whose cross-section is a magnetic island or bubble). 
    (c) Flux-rope  grows in size and erupts, carrying cool plasma upwards. 
    (d) Close-up of the reconnection region where a hot jet is accelerated.} 
    \label{fig10}
\end{figure}

\subsection{Phase 2 of cancellation}
\label{sec4.2}

There has been a debate on the actual process of flux cancellation in the photosphere, dating back to \citet*{zwaan87} and \citet*{priest87} and others, as summarised in, for example,  \citet*{priest94b}. One suggestion was that it represents pure flux submergence (without reconnection nearby) and another was that it is caused by magnetic reconnection. If reconnection occurs at the photosphere, then the photospheric cancellation is occurring in the reconnection site. If, however, reconnection occurs just above the photosphere, then cancellation represents the submergence of inverted U-loops retracting down through the photosphere after having been reconnected. The argument for  reconnection in either location, which we favour here, is that it would then naturally produce the energy release that is often observed in the form of heating and plasma acceleration. 

As can be seen in Fig. \ref{fig10}a, while the two polarities approach and eventually come in contact, the field above the polarity inversion line becomes non-potential, and will form another localised current sheet (different from the one discussed in Phase 1), which extends upwards from the photosphere or above. In a vertical plane through the cancellation process (panel b), a magnetic bubble or island is naturally produced by reconnection at or just above the photosphere. Indeed, this naturally carries cool plasma from the photosphere and chromosphere upwards, as has been proposed by, for example, \citet{sterling15,sterling16a,sterling16b,sterling20a}. 
The cool plasma they dub a `mini-filament'.  If there is an extra component of magnetic field out of the plane, as is usually the case, then the magnetic island is just the cross-section of a magnetic flux rope or a small sheared arcade.
We note that the initiation of Phase 2 of the cancellation does not have to wait until Phase 1 ends. The cancellation process starts with Phase 1, but Phase 2 can occur while Phase 1 is still on-going. The timing between the two phases will depend on the magnetic configuration, area, flux content, and distance between the two cancelling polarities.

The physical properties of flux cancellation during Phase 1 with reconnection in the atmosphere have been estimated in \cite{priest18} and \cite{syntelis19}, and so we now estimate the corresponding properties during Phase 2 with reconnection in the photosphere as follows. They vary hugely, depending on the size and field strength of the flux and of the length of the current sheet. The  sheet length depends on the nature of reconnection. For Sweet-Parker reconnection, the length of the current sheet would be $ l=\eta v_A/v_i^2$, where values of the magnetic diffusivity $\eta=10^4$ m$^2$/sec, Alfv\'en speed $1-10$ km s$^{-1}$ and inflow speed $v_i=10^3$ m/sec would give a length of only $L=0.01-0.1$ Mm, so that the released energy would  generally be too small to explain the observations.

For fast reconnection, on the other hand, the energy release is much larger since
the current sheet now refers not just to a tiny Sweet-Parker sheet, but also to the bifurcated sheet including the slow shock waves for Petschek reconnection, or to a turbulent current sheet for impulsive bursty reconnection or a collisionless Hall sheet, as discussed in \citet*{syntelis19}. If the sheet extends up to a height of, say, 1 Mm in the atmosphere, then the energy release is sufficient, as the following estimates show, to account for microflares and subflares and on much smaller scales for nanoflares.
Thus, for the various kinds of fast reconnection, most of the energy is not liberated in the central Sweet-Parker sheet but in the bifurcated or turbulent part of the sheet.  Also, the observed decline in energy release as time proceeds could be due to a decline in field strength $B_i$ and or cancellation speed $v_i$.

A magnetic flux tube of radius $R_0$ and field strength $B_0$ has a flux $F=\pi R_0^2B_0$, which may be written as
\begin{equation}
F=3\ R_1^2\ B_{100} \times 10^{18}\ {\rm Mx},
\label{eq4.24}
\end{equation}
where $R_1$ is the radius in units of Mm and $B_{100}$ is the magnetic field in units of $100$ G. Thus, for example, a magnetic fragment of radius $1$ Mm and field of $1$ kG has a flux of $3\times 10^{19}$ Mx, whereas if the radius is only $50$ km, then the flux is $7.5\times 10^{16}$ Mx.

The velocity ($v_i$) and duration ($\tau$) of the cancellation of tubes of radius $R_0$ are related by $v_i=2R_0/\tau$ or
\begin{equation}
v_i=\frac{2R_1}{\tau_1},
\label{eq4.26}
\end{equation}
where $\tau_1$ is measured in units of 1000 sec, and so for a radius of 0.7 Mm and a duration of 3000 sec, the cancellation speed would be $v_i=0.5$ km s$^{-1}$.

The energy released during cancellation may be estimated in two ways as follows. The first estimate is to consider two magnetic flux tubes of radius $R$, and field strength $B$, each with a magnetic energy of $\pi R^2  B^2/(4 \pi)$ per unit length.
If two such tubes cancel over a length $L$, the energy released is
\begin{equation}
W=5 \ R_1^2\ L_1\ B_{100}^2  \times 10^{27}\ {\rm erg},
\label{eq4.25}
\end{equation}
where $L_1$ and $R_1$ are measured in Mm and $B_{100}$ in hundreds of Gauss.

The second estimate is to consider the rate of release of energy in a sheet of width $2R_0$ and height $L$, as  given by
\begin{equation}
\frac{dW}{dt} = \frac{v_iB_i^2}{4 \pi} 2R_0 L,
\label{eq4.27}
\end{equation}
where $v_i=2R_0/\tau$, and so, during a time $\tau$, an energy
\begin{equation}
W=2\frac{B_i^2}{4\pi} L(2R_0)^2=6 \ R_1^2\ L_1\ B_{100}^2  \times 10^{27}\ {\rm erg}
\nonumber
\end{equation}
is released, which is of the same form as Eq. (\ref{eq4.25}) and depends crucially on the length of the current sheet.

We then consider first two tiny intense flux tubes of radius $50$ km with fields of $1$ kG and a sheet length of $1$ km. If the cancellation speed is $1$ km s$^{-1}$, it will produce an energy of $10^{24}$ erg over $100$ sec, which is appropriate for a nanoflare.
On the other hand tubes of radius $0.5$ Mm with fields of $100$ G and a length of $1$ Mm  cancelling at a speed $1$ km s$^{-1}$ yield an energy of $10^{27}$ erg over $10^3$ sec appropriate for a microflare, whereas tubes of radius $1$ Mm with fields of $500$ G and a length $L= 1$ Mm give an energy of $10^{29}$ erg typical of a subflare over $200$ sec.  Also, we note that if the larger flux elements consist of many finer intense flux tubes with persistent flux cancellation, or if the cancellation occurs in fits and starts, then the total energy release may take place as a series of nanoflares or microflares over an extended time of hundreds or thousands of seconds, as reported in some observations and simulations of flux cancellation such as \cite{peter19} and \cite{park20}. 

\section{Conclusions}

Magnetic flux cancellation was previously realised to be important in heating tiny regions of the solar atmosphere, namely, X-ray bright points. However, the  Sunrise observations have transformed our appreciation of its significance and shown flux cancellation to be very much more widespread, and therefore potentially to be the dominant factor in heating the atmosphere and accelerating various kinds of jets in different parts of the solar atmosphere.
The aim of the present paper has been to further develop, in several directions, the basic theory for such energy release driven by flux cancellation.

The first direction was a technical one, namely, to determine how the previous simple theory of reconnection at a Cartesian current sheet in 2D can be set up without using complex variable theory and how it can be extended to 3D. For an axisymmetric toroidal current sheet, we have found how the large-scale curvature decreases the field outside the torus. Then we applied the theory to flux cancellation between two magnetic fragments, where we realised there are two stages, namely, (i) Phase 1, during which reconnection occurs at a separator that first moves up and then descends back to the photosphere and (ii) Phase 2, during which reconnection occurs in or just above the photosphere between the two cancelling regions.

Future possible developments include detailed computational experiments that can produce more realistic models for the process and can validate the basic theory that we have proposed here. In addition, we have focused here on conceptually the simplest building block of the theory, namely, the elementary interaction of two magnetic fragments, but in future it will be possible to apply the theory to a variety of more complex and realistic geometries and flux systems.

\begin{acknowledgements}
ERP is grateful for helpful suggestions and hospitality to Pradeep Chitta, Hardi Peter, Sami Solanki and other friends in MPS G\"ottingen, where this research was initiated. P.S. acknowledge support by the ERC synergy grant ``The Whole Sun''. 
The authors are most grateful for a thorough and insightful referee report that has substantially improved the paper.
\\

\end{acknowledgements}

\bibliographystyle{aa}
\bibliography{bibliography}

\end{document}